\documentclass[11pt,twoside]{article}

\usepackage{asp2014}

\aspSuppressVolSlug
\resetcounters

\begin{document}

\title{JAvaScript Multimodal INformation Explorer}

\author{Fenja Schweder,$^{1,2}$, Sebastian Trujillo-Gomez$^2$, and Kai Polsterer,$^2$}
\affil{$^1$University of Bremen, Bremen, Germany; \email{fen\_sch@uni-bremen.de}}
\affil{$^2$HITS gGmbH, Heidelberg, Germany}

\paperauthor{Fenja~Schweder}{fen\_sch@uni-bremen.de}{0009-0001-8069-9952}{University of Bremen}{Digital Media Lab}{Bremen}{Bremen}{28359}{Germany}
\paperauthor{Trujillo-Gomez~Sebastian}{sebastian.trujillo-gomez@h-its.org}{}{HITS gGmbH}{AIN}{Heidelberg}{Baden-Württemberg}{69118}{Germany}
\paperauthor{Polsterer~Kai}{kai.polsterer@h-its.org}{}{HITS gGmbH}{AIN}{Heidelberg}{Baden-Württemberg}{69118}{Germany}



\begin{abstract}
Astronomical data is rich in volume, information and facets. Although this offers multiple research perspectives, processing the data remains a challenge. Infrastructures for analyzing, inspecting, exploring and communicating with data are mandatory.
To address this issue, we introduce Jasmine, the JAvaScript Multimodal INformation Explorer. Jasmine allows users to open different data viewer modals that show a specific data point from a set. The viewer currently supports image data, as well as point cloud objects. Users can decide on which information about the data point they like to have displayed. Point clouds are interactive and allow for zooming, tossing, and turning. Picking a data point is enabled by providing a structured view of the set, arranged by a key property. This arrangement is achieved by autoencoding. 
\end{abstract}



\section{Introduction}
Large surveys and extensive simulations sustain astronomers with information that is not only large in volume but also covers various aspects, fields, and dimensions. The effort in analyzing and processing the data is a remaining issue extensively described by \cite{zhang15} and is still under review (\citet{faarique23}). This not only raises the expectation in algorithms to process and analyze the data but also requires infrastructures to apply them and work with the results.

Visualization and interactivity play a key role for the mentioned infrastructures. They have been addressed in earlier proposals for interactive exploration frameworks. \citet{sciacca14} introduced a platform to facilitate access, visualization, and exploration of big astronomical data. Another example is \textit{Astronomaly} by \cite{lochner21}, a web application detecting outliers in big data sets. \textit{UltraPINK} \citep{kollasch24} visualizes common structures in a data set as Self-Organizing Kohonen Maps. 

The mentioned solutions address the difficulty of providing a structured overview among a large data set. However, an additional challenge is displaying individual data points with multiple fields. A single visualization of a multidimensional data point is not capable of displaying all contained information without being overloaded. In addition, some aspects require different visualization techniques from others, due to their dimensionality or behavior. Desirable is a data inspector that offers the ability to switch between dimensions of interest. 

\section{The Concept of multimodal data exploration}
In this work, we introduce the \textit{JAvaScript Multimodal INformation Explorer}. JASMINE is a prototypical web application that enables inspecting and interacting with multidimensional data points from a big data set. The key feature of JASMINE is the gathering and comparing of different views on an astronomic object with multiple modals. In the following, we explain the underlying concept and how to implement the concept to address the requirements of astronomical data.

\subsection{Designing a big data browser}
The book \textit{The Craft of Information Visualization} by \cite{benderson03} describes useful guidelines on how to create a user-friendly and comprehensible scientific visualization. Included in this work, \cite{plaisant03} provide comprehensive strategies for designing information browsers. Our approach of a multidimensional data explorer follows their taxonomy of multiple-view browsers. 

Based on Shneiderman's mantra \textit{Overview first, zoom and filter, then details-on-demand} \citep{shneiderman03}, the launch window shows an overview among the whole data set. Selecting a data point will open an individual modal for the point where further and more detailed inspection is possible. Communication between modals is unidirectional from the overview window to the detail windows. By that means, interaction with the overview window may change the view in the detail window, but not vice versa. 

\cite{plaisant03} recomment multi-window browsers with window placement and alignment strategies. Leaving the window management to the user also leaves them with additional workload. On the other hand, it is difficult to find an appropriate strategy that fulfills the user's needs for multiple cases. This is especially an issue with information browsers for explorative research, where the task focus may not be clear at first.

\subsection{Massive and multivariate data}
 Since we expect the amount of data to be extremely large, an appropriate representation, if necessary aided by machine learning, is or can be chosen. \citet{kollasch22} explain briefly how to apply techniques such as clustering, regression, and dimensionality reduction to create an interactive visualization of a big data set. The respective techniques are capable of creating an overview among the data, where strong features are outlined. The relations between the representation and the real data are preserved.
 
 Showing astronomical objects with multiple fields in a single visualization is not possible without the visualization being overloaded. For this type of multivariate data, we designed a data type allowing one to store each field as individually accessible. Each data point from the original set is saved in this \textit{data cube} format. Each side of the cube contains a different field that belongs to the underlying astronomical phenomenon. Data cubes can be nested to ensure a hierarchy among the properties, meaning that each side can have an arbitrarily large set of subsides. Users have the option to select a cube side to be shown in the detail window. The chosen side has to of the lowest hierarchy. Thanks to the multimodal approach, the cube sides and therefore the data fields can be directly compared.

\section{Prototype}
We sketch a prototypical implementation of the introduced concepts in JavaScript, together with a use-case example of a massive, multivariate data set. The result is a browser-driven web application that runs client-side. Against the recommendation of \cite{plaisant03}, we did not implement a window management strategy, but instead use web browser windows as modals. Due to the variety, finding an appropriate strategy to automatically organize, align, and resize the windows requires specific attention and will be addressed in the future. Instead, the user is now free to design the browsing windows according to their requirements.

\subsection{Example data and preparation}
As a case of browsing extremely large multivariate data sets, we take snapshots from the \textit{Illustris} project \citep{nelson18} as an example. Illustris TNG is a set of large cosmological simulations that predict the formation and evolution of galaxies including many complex physical processes. The simulations model representative cosmological volumes across the history of the universe, following the evolution of multiple thousands galaxies. Each snapshot contains different types of particles. Each particle field has in turn multiple physical properties.

We took a subsample of 1000 galaxies that are provided with a stellar mock image. To generate an explorable surface, we use \textit{Spherinator \& HiPSter} framework \citep{polsterer24}. The \textit{Spherinator} learns the intrinsic morphological features of galaxies from the mock images. It projects an arbitrarily large dataset using an interactive hierarchical spherical representation. The learned latent representation of galaxy morphological space is interpretable. The 3D positions in latent space reflect the morphological properties of each galaxy. To explore them, the \textit{HiPSter} creates a hierarchical view as a HiPS tiling that can be examined with \textit{Aladin Lite} \citep{baumann22}.

The last preparation step is organizing each galaxy from the sample as a stacked data cube with a side for each particle type and a subside for each data field, respectively. To do this on the fly with the HiPS tile generation, we apply light modifications to the \textit{HiPSter} source code.

\subsection{Dataset-driven methods}
The interaction methods on the global dataset level are directly inherited from \textit{Aladin Lite}. Zooming in increases the number of HealPiX cells and therefore the number of galaxies projected to the sphere surface. A dropdown menu allows one to change the view layer between the latent space reconstructions and the original images. Furthermore, it is possible to superimpose the catalog coordinates for all the objects and inspect the galaxy metadata interactively with access to the individual images and all the simulation data for that object directly through the TNG API.

Right clicking on a cell opens a detail window for the respective galaxy. The prototype can be configured to show either one detail window at a time or multiple windows of the same or different galaxies. The user can arrange, resize, and remove modals as required.

\subsection{Datapoint-driven methods}
The detail modal shows a selected subside of the data cube. By default, the mock image used to train the morphologic representations is shown. Since Illustris TNG provides multiple particle clouds representing the gas, stellar, or dark-matter distribution, the user has the option to examine the 3D structures of the galaxy. JASMINE allows interacting with particle clouds via rotating and zooming. Color-coding shows the field values for the active particle clouds. A set of radio buttons lets the user determine which field should be shown. To make the difference clear, each field as a different color scale.

\section{Conclusion and Outlook}
JASMINE provides basic functionalities to explore a very large data set from a global to a detailed view point. The prototype shows the potential that comes with a multimodal data explorer and offers multiple ideas for future work. A window managing strategy suitable for mutable data would drastically increase the user experience. Bidirectional communication between detail windows and global view offers new possibilities to explore the data. Lastly, we plan for a generalized interface that allows us to attach JASMINE to any kind of data processing pipeline as an abstract data browser.

\bibliography{P406}  


\end{document}